\documentclass[prd, preprint, nofootinbib, showpacs]{revtex4}
\usepackage{epsf,epsfig,subfigure,graphicx,amsmath,amssymb}
\usepackage{amsfonts}
\usepackage{latexsym}
\usepackage{color}

\newcommand{\dis}[1]{\begin{equation}\begin{split}#1\end{split}\end{equation}}
\newcommand{\be}{\begin{equation}}
\newcommand{\ee}{\end{equation}}
\def\bea{\begin{eqnarray}}
\def\eea{\end{eqnarray}}

\newcommand{\VEV}[1]{\langle #1 \rangle}
\newcommand\tev{\,{\rm TeV}}
\newcommand\gev{\,{\rm GeV}}

\begin{document}

%\begin{flushright}
%\end{flushright}

\title{
Vector Higgs-portal dark matter and Fermi-LAT gamma ray line \\[5mm]
}

\author{Ki-Young Choi}
\email{kiyoung.choi@apctp.org}
 \affiliation{
 Asia Pacific Center for Theoretical Physics, Pohang, Gyeongbuk 790-784, Republic of Korea and \\
 Department of Physics, POSTECH, Pohang, Gyeongbuk 790-784, Republic of Korea}

\author{Hyun Min Lee}
\email{hyun.min.lee@kias.re.kr}
 \affiliation{School of Physics, KIAS, Seoul 130-722, Republic of Korea}

\author{Osamu Seto}
 \email{seto@physics.umn.edu}
\affiliation{
 Department of Life Science and Technology,
 Hokkai-Gakuen University,
 Sapporo 062-8605, Japan}

%\date{\today}
%

%%%%%%%%%%%%%%%%%%%%%%
\begin{abstract}
%%%%%%%%%%%%%%%%%%%%%%
%
We propose a renormalizable model for vector dark matter with extra $U(1)$ gauge symmetry, which is broken by the VEV of a complex singlet scalar. 
When the singlet scalar has a quartic coupling to a heavy charged scalar, the resonance effect enhances the annihilation cross section of  vector dark matter into two photons such that Fermi-gamma ray line at  about $130 \gev$ is obtained. In the presence of a tiny mixing between the singlet scalar and the Standard Model Higgs doublet, the relic density is determined dominantly by $WW/ZZ$ and two-photon channels near the resonance pole of the singlet scalar.
We also show that various phenomenological bounds coming from the Higgs to diphoton decay rate, precision data and collider searches for the charged scalar and vacuum stability are satisfied in the model.
%
%%%%%%%%%%%%%%%%%%%%%%
\end{abstract}
%%%%%%%%%%%%%%%%%%%%%%

\pacs{95.35.+d, 12.60.Cn, 98.70.Vc}

\preprint{KIAS-P13018, HGU-CAP-020, APCTP Pre2013-005} 

\vspace*{1cm}
\maketitle

%==================================%
%          Main body               %
%==================================%

%%%%%%%%%%%%%%%%%%%%%%%
\section{Introduction}
\label{sec: introduction}
%%%%%%%%%%%%%%%%%%%%%%%

The existence of dark matter (DM)~\cite{DMreview} provides one of the strong motivations
 to seek for physics beyond the Standard Model (SM).
Weakly interacting massive particles (WIMPs) have been the promising candidate for dark matter,
 that are assumed to have weak scale interactions with the SM particles and a weak scale mass. 
However, the property and identity of WIMP dark matter remains unknown and
 is one of the big puzzles in particle physics and cosmology. 

Recently, it has been shown that gamma ray line spectrum exists at $E_\gamma=130 \gev$ using Fermi-LAT data~\cite{Bringmann:2012vr,Weniger:2012tx}. 
The signature has been independently confirmed by other groups~\cite{Tempel,Su:2012ft} and officially
 investigated by the Fermi-LAT collaboration but with the peak being shifted to $E_\gamma =135$ GeV~\cite{Bloom:2013mwa}. 
There are some possible explanations such as mono-energetic pulsar winds~\cite{Aharonian:2012cs},
 Fermi Bubble~\cite{Su:2010qj,Profumo:2012tr}
 or instrumental effects~\cite{Whiteson:2012hr,Finkbeiner:2012ez,Whiteson:2013cs}
 including Earth limb signal~\cite{Hektor:2012ev,Bloom:2013mwa}. 
The Fermi-LAT collaboration~\cite{Ackermann:2012qk}
 and the H.E.S.S collaboration~\cite{Abramowski:2013ax} have reported only the upper bound
 on the annihilation cross section of WIMPs, that is compatible with the Fermi gamma-ray line, so the dark matter interpretation of the Fermi-LAT line signature seems plausible.

Dark matter is neutral and thus cannot annihilate into photons at tree-level. 
The generation of photons must happen via the loops of charged particles to which dark matter is directly or indirectly coupled, so  the annihilation cross section of dark matter into photons
 is much suppressed as compared to other tree-level annihilation channels. 
Thus, in order to realize a large branching fraction of the annihilation cross section into photons,
 we need to rely on a large coupling to new charged particles running in loops~\cite{Cline,Choi:2012ap,Tulin:2012uq}
 or a resonance pole of the mediator particle between dark matter and photons~\cite{resonance,resonance2}.

The extra $U(1)$ gauge symmetry is one of the simplest extensions of the SM.
In this paper, we propose a renormalizable model of vector dark matter
 in which the extra $U(1)_X$ gauge boson couples to the SM particles
 through the mixing between the complex singlet scalar, which is responsible for $U(1)_X$ breakdown, and the SM Higgs doublet. 
When there is a quartic coupling between the singlet scalar and a heavy charged scalar, dark matter
 can annihilate into a photon pair with sizable branching fraction,
 provided that the mixing between the singlet scalar and the SM Higgs boson is small enough. 
The annihilation cross section of dark matter into a photon pair is enhanced near the resonance pole of the singlet-like scalar mediator  to be consistent with Fermi gamma-ray line. For a tiny mixing between the neutral scalars, the thermal relic density can be determined dominantly by annihilation channels into $WW$, $ZZ$ and a photon pair.
We discuss various phenomenological implications of the model, e.g., Higgs to diphoton rate, electroweak precision measurements, collider constraints on the charged scalar, and the vacuum stability bound of the scalar potential.

We note that in most of other previous works on vector WIMP, a vector boson was introduced as a Proca field~\cite{vdm,vdm2,farzan} or a scalar sector is described by the nonlinear
 sigma model~\cite{Abe:2012hb}.
In our model, vector dark matter is realized as a gauge boson of extra $U(1)$ symmetry when it is broken spontaneously by the VEV of the hidden Higgs field at the renormalizable level \cite{vdm2,pko}.

The paper is organized as follows. 
Beginning with the introduction of our model,
in Sec.~\ref{sec: dark matter},
 we calculate the annihilation cross section of the $U(1)_X$ gauge boson dark matter
 and show that the desired thermal relic density and a large annihilation cross section into two monochromatic photons
 can be realized for a consistent set of parameters.
Then, we study the effect of the charged scalar on the Higgs diphoton decay rate in Sec.~\ref{sec: Higgs diphoton},  and  various experimental constraints on the charged scalar in Sec.~\ref{sec: constraints on charged scalar}.
In Sec.~\ref{sec: vacuum stability}, we study the running of the couplings
with the modified renormalization group equations (RGEs) due to additional interactions and consequently show how the stability of the Higgs potential is improved.
We summarize our results in Sec.~\ref{sec: conclusion}. 
There are two appendices summarizing the gauge and scalar interaction vertices and RG equations, respectively.

%%%%%%%%%%%%%%%%%%%%%%%
\section{The model}
\label{sec:model}
%%%%%%%%%%%%%%%%%%%%%%%

We consider a simple model of vector WIMP dark matter which couples to the SM particles through
 Higgs portal interactions.
The gauge sector of the model
 is based on the $SU(3)_C \times SU(2)_L \times U(1)_Y \times U(1)_X$ gauge group.
The extra $U(1)_X$ gauge symmetry is spontaneously broken by
 the vacuum expectation value (VEV) of a complex scalar $S_1$.
For a minimal extension of the SM \footnote{The model can be extended with a large multiplet containing a charged scalar such as extra Higgs doublet or Higgs triplet \cite{triplet}, etc.}, we introduce an $SU(2)$ singlet charged scalar $S_2^+$, which carries $Y=1$ but is neutral under $U(1)_X$. We assume that all the SM particles including the Higgs doublet are neutral under the $U(1)_X$.

The model has a $Z_2$ symmetry under which  $S_1\rightarrow S_1^*$ and $X_\mu\rightarrow - X_\mu$, which guarantees the stability of vector dark matter $X_{\mu}$.
The Lagrangian of the model is
\begin{eqnarray}
{\cal L} &=&  -\frac{1}{4}F_{\mu\nu} F^{\mu\nu} 
 + |D_\mu S_1|^2+|D_\mu S_2|^2-V(\Phi,S_1,S_2 )
 + f_{ij} L_i C \cdot L_j S_2^+ ,
\label{lagrangian}
\end{eqnarray}
where $F_{\mu\nu}=\partial_\mu X_\nu-\partial_\nu X_\mu$, and the covariant derivatives are $D_\mu S_1 =( \partial_{\mu} -i g_X X_{\mu})S_1$ and $D_\mu S_2=(\partial_\mu -i g' B_\mu)S_2$, with respect to
 $U(1)_X$ and $U(1)_Y$ gauge symmetry, respectively. 
After the electroweak symmetry breaking, $D_\mu S_2$ is reduced to the covariant derivative with respect to $U(1)_{\rm em}$ symmetry. The scalar potential of the SM Higgs doublet $\Phi$ and complex scalar fields $S_1$ and $S_2$ is given by
\begin{eqnarray}
V(\Phi,S_1,S_2 ) &= & \mu^2_1 |\Phi|^2 + \mu^2_2 |S_1|^2 + \mu^2_3 |S_2|^2 +\frac{1}{2}\lambda_1 |\Phi|^4+\frac{1}{2}\lambda_2 |S_1|^4+\frac{1}{2}\lambda_3 |S_2|^4
\nonumber\\
 && +\lambda_4 |\Phi|^2|S_1|^2 +\lambda_5 |\Phi|^2|S_2|^2
 +\lambda_6 |S_1|^2|S_2|^2.
\label{scalar potential}
\end{eqnarray}
Here the $\lambda_4$ coupling is relevant for the mixing between $\Phi$ and $S_1$, after the breaking of $U(1)_X$ and electroweak symmetry. 
The dominant annihilation of vector dark matter occurs through this coupling
so that the correct relic density can obtained. 
The coupling $\lambda_6$ connects the vector dark matter to the charged particle and enhances the photon emission. 
The coupling $\lambda_5$ in combination of the charged scalar mass can be constrained
 by the branching ratio of Higgs boson decay into two photons at the Large Hadron Collider (LHC). 
 
In the last term of the Lagrangian (\ref{lagrangian}), $L_i$ is the SM lepton doublet with flavor index $i=1,2,3$, $C$ is the charge-conjugation operator and dot denotes the $SU(2)$ anti-symmetric product. 
This term induces the decay of heavy charged scalar $S_2^\pm\rightarrow l_i^\pm + \bar{\nu}_j $. 
The experimental constraints on the lepton flavor violating term will be discussed in
 the Sec~\ref{sec: constraints on charged scalar}. 
This model can be extended to a Type-II seesaw model where the charged scalar $S_2$ is embedded
 into a triplet Higgs field with $Y=2$~\cite{tripletH}. 
In this case, the lepton couplings of the charged scalar would be small in order to explain neutrino masses.

At the vacuum with a nonvanishing singlet VEV, $\langle S_1\rangle =v_S/\sqrt{2}$, the $U(1)_X$ gauge symmetry is broken so
 the gauge boson $X_{\mu}$ acquires mass,
\begin{equation}
 M_X^2 = g_X^2 v_S^2 .
\label{mass:X}
\end{equation}
We expand $\Phi$ and $S_1$ fields in unitary gauge, around the electroweak vacuum with $\langle \Phi\rangle=v/\sqrt{2}$ and $\langle S_1\rangle =v_S/\sqrt{2}$, as
\begin{eqnarray}
\Phi &=& \left( 
            \begin{array}{c}
            0 \\
            \frac{1}{\sqrt{2}}(v + \phi)  
            \end{array}   \right) ,  \\
S_1 &=&  \frac{1}{\sqrt{2}}(v_S + \phi_S) ,
\end{eqnarray}
with $v\simeq246\gev$.
Then, two scalar modes $\phi$ and $\phi_S$ in general mix so
 the mass eigenstates $h$ and $H$ are given in terms of the mixing angle $\alpha$ as
 \dis{
  h=\cos\alpha \, \phi -\sin \alpha \, \phi_S,
 &\qquad H=\sin\alpha \, \phi + \cos \alpha \, \phi_S.
 }
The Higgs mixing essentially depends on $\lambda_4$ through
\begin{eqnarray}
\tan2\alpha = \frac{2 \lambda_4 v v_S}{ \lambda_1 v^2-\lambda_2 v^2_S},
\end{eqnarray}
and the mass eigenvalues are
\begin{eqnarray}
M^2_{h,H}= \lambda_1 v^2+\lambda_2 v^2_S\mp \sqrt{(\lambda_2 v^2_S-\lambda_1 v^2)^2+4\lambda^2_4 v^2 v^2_S}\,.
\end{eqnarray}
For a small mixing angle $\alpha$, we can regard $ h \simeq \phi $ as being SM Higgs-like and $H \simeq \phi_S$ as being singlet-like,
 and the mass eigenvalues are $M_h\simeq \lambda_1 v^2$ and $M^2_H\simeq \lambda_2 v^2_S$.
The gauge and scalar interactions of the Higgs fields are listed in Appendix A.

The absolute stability of the electroweak vacuum gives rise to the following conditions on the quartic couplings in the scalar potential~\cite{kannike},
\bea
\lambda_1 >0,\quad \lambda_2 >0,\quad \lambda_3 >0,  \nonumber \\
\lambda_{12}\equiv\lambda_4+\sqrt{\lambda_1 \lambda_2}>0,\\
 \lambda_{13}\equiv\lambda_5+\sqrt{\lambda_1\lambda_3}>0, \nonumber \\
 \lambda_{23}\equiv\lambda_6+\sqrt{\lambda_2\lambda_3}>0,\nonumber
\eea
and  
\be
\lambda_{123}\equiv \frac{1}{2}\sqrt{\lambda_1\lambda_2\lambda_3}+\lambda_6\sqrt{\lambda_1}+\lambda_5\sqrt{\lambda_2}+\lambda_4\sqrt{\lambda_3}+\sqrt{\lambda_{12}\lambda_{23}\lambda_{13}}>0.
\ee
Throughout the work, we also impose perturbativity conditions until the Planck scale as
\bea
|\lambda_i| <4 \pi.
\eea

%%%%%%%%%%%%%%%%%%%%%%%
\section{Relic density and Fermi gamma-ray lines}
\label{sec: dark matter}
%%%%%%%%%%%%%%%%%%%%%%%

In this section, we compute the DM annihilation cross sections and discuss the constraints of the relic density and the Fermi gamma-ray line in the model.

\subsection{Thermal relic abundance}
\label{subsec: thermal}
\begin{figure}[!t]
  \begin{center}
  \begin{tabular}{c}
   \includegraphics[width=0.5\textwidth]{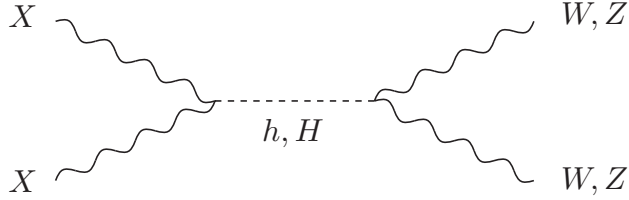}
     \end{tabular}
  \end{center}
  \caption{Feynman diagram for annihilations of vector dark matter into $WW, ZZ$ at tree level.}
  \label{fig:XX2WW}
\end{figure}

%%%%%%%%%%%%%%%%%%%%%%%
The thermal relic abundance of the vector WIMP dark matter $X$ is estimated 
 by integrating the following Boltzmann equation for
 the dark matter number density $n_X$
 in the early Universe~\cite{Boltzmann,Gondolo:1990dk},
\begin{equation}
\frac{dn_X}{dt} + 3H n_X =
-\langle \sigma v \rangle \Big(n^2_X - (n^{\rm EQ}_X)^2\Big),
\end{equation}
 where $H$, $\langle\sigma v\rangle$, and $n^{\rm EQ}_X$ denote the Hubble parameter,
 the thermal averaged annihilation cross section times relative velocity, 
 and the dark matter number density at thermal equilibrium, respectively. 
$X$ dominantly annihilates into $W$ and $Z$ boson pairs
 through the s-channel exchange of the Higgs bosons $h$ and $H$, as shown in Fig.~\ref{fig:XX2WW}. 
The magnitude of those annihilation cross sections is proportional to $(\sin\alpha\cos\alpha)^2$
 and thus scaled by the mixing angle between 
 Higgs bosons.
A larger (smaller) annihilation cross section is realized for 
 a larger (smaller) $\sin\alpha$.
The velocity times DM annihilation cross sections into $W,Z$ boson pairs before thermal average are given by
\dis{
(\sigma v)_{WW}(s) =&\frac{1}{18 \pi s} 
 \left| 
  \frac{g_{XXh}g_{hWW}}{s-M_h^2+i M_h \Gamma_h} 
  + 
  \frac{g_{XXH} g_{HWW}}{s-M_H^2+i M_H \Gamma_H }  \right|^2 \\
  &\times   \left\{ 1 +\frac{1}{2 M_X^4}\left(\frac{s}{2} - M_X^2\right)^2 \right\} \left\{ 1 +\frac{1}{2 M_W^4}\left(\frac{s}{2} - M_W^2\right)^2 \right\} \sqrt{1-\frac{4M_W^2}{s}},
}
and
\dis{
(\sigma v)_{ZZ}(s) =&\frac{1}{36 \pi s} 
 \left| 
  \frac{g_{XXh}g_{hZZ}}{s-M_h^2+i M_h \Gamma_h} 
  +   \frac{g_{XXH} g_{HZZ}}{s-M_H^2+i M_H \Gamma_H }  \right|^2 \\
  &\times  \left\{ 1 +\frac{1}{2 M_X^4}\left(\frac{s}{2} - M_X^2\right)^2 \right\}  \left\{ 1 +\frac{1}{2 M_Z^4}\left(\frac{s}{2} - M_Z^2\right)^2 \right\}  \sqrt{1-\frac{4M_Z^2}{s}},
}
where $s$ is the total energy at the center of mass frame and each couplings are given by
\dis{
g_{XXh}&=-2g_X^2 v_S \sin\alpha, \qquad g_{XXH}= 2g_X^2 v_S\cos\alpha,\\
g_{hWW}& = \frac12 g_2^2 v \cos\alpha , \qquad g_{HWW} = \frac12 g_2^2 v \sin\alpha,\\
 g_{hZZ}& = \frac{ g_2^2}{2\cos\theta_W^2} v \cos\alpha , \qquad g_{HZZ} = \frac{ g_2^2}{2\cos\theta_W^2} v \sin\alpha.
 }

Moreover, the corresponding expression for the DM annihilation into $hh$ is also given by
\dis{
(\sigma v)_{hh}(s) =&\frac{1}{72 \pi s} 
 \left| 
  \frac{g_{XXh}g_{hhh}}{s-M_h^2+i M_h \Gamma_h} 
  +   \frac{g_{XXH} g_{Hhh}}{s-M_H^2+i M_H \Gamma_H }  \right|^2 \\
  &\times  \left\{ 1 +\frac{1}{2 M_X^4}\left(\frac{s}{2} - M_X^2\right)^2 \right\}    \sqrt{1-\frac{4M_h^2}{s}},
}
where
\bea
g_{hhh}&=& 3v(\lambda_1 \cos^3\alpha+\lambda_4 \sin^2\alpha \cos\alpha)-3v_S(\lambda_2\sin^3\alpha +\lambda_4 \sin\alpha \cos^2\alpha), \\
g_{Hhh}&=&\frac{1}{4}v\Big(\sin\alpha\,(3\lambda_1+\lambda_4)+ 3 \sin3\alpha\, (\lambda_1-\lambda_4)\Big) \nonumber \\
&&+\frac{1}{4}v_S\Big(\cos\alpha\,(3\lambda_2+\lambda_4)+3\cos3\alpha\, (\lambda_4-\lambda_2)\Big).
\eea
However, we find that the annihilation cross section of the $hh$ channel is numerically smaller than those of the $WW,ZZ$ channels by order of magnitude, which has the polarization sum over the final states.
The other annihilation channels with $hH$ and $HH$ final states are kinematically forbidden near the resonance, $M_H\sim 2 M_X$.

\begin{figure}[!t]
  \begin{center}
  \begin{tabular}{ccc}
   \includegraphics[width=0.4\textwidth]{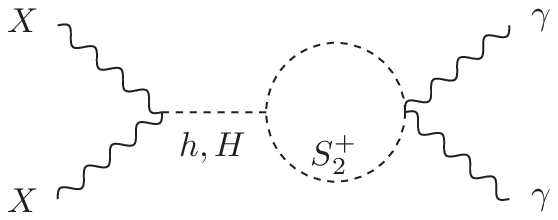}
&\qquad \qquad\qquad &
   \includegraphics[width=0.4\textwidth]{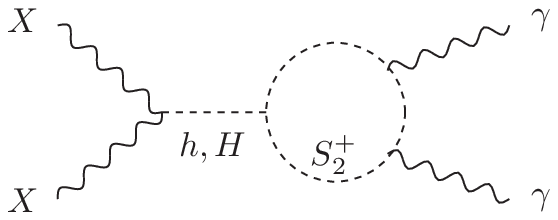}
     \end{tabular}
  \end{center}
  \caption{Feynman diagrams for the annihilation of vector dark matter into a photon pair at one-loop.}
  \label{fig:XX2gg}
\end{figure}

Finally, when there is a sizable quartic coupling between the singlet scalar and the heavy charged scalar $S_2$, dark matter can annihilate into a photon pair, due to loops with the charged scalar $S^+_2$ as shown in Fig.~\ref{fig:XX2gg}. The velocity times DM annihilation cross section into a photon pair before thermal average is expressed by
\begin{eqnarray}
(\sigma v)_{\gamma\gamma}(s)
 &=& \frac{ \alpha_{\rm em}^2}{96 \pi^3 s}  
  \left|\sum_{H_i = h, H} \frac{g_{XX H_i} g_{H_i S_2^+S_2^-} }{s-M_{H_i}^2+i M_{H_i}\Gamma_{H_i}}\right|^2
  \left| 1 - \frac{ M_{S_2^+}^2}{M_X^2} f\left(\frac{ M_{S_2^+}^2}{M_X^2}\right) \right|^2 \nonumber \\
&& \times  \left\{ 1 +\frac{1}{2 M_X^4}\left(\frac{s}{2} - M_X^2\right)^2 \right\} ,
\label{sigmav:twogamma}
\end{eqnarray}
with
\begin{eqnarray}
 f(\tau) = \Big[\sin^{-1}(1/\sqrt{\tau})\Big]^2 .
\end{eqnarray}
Depending on the mixing angle $\alpha$, the DM annihilation into a photon pair can have a sizable branching fraction, as will be illustrated in the later subsection.
 
The heavy Higgs boson mainly decays into $XX$, and also into $W,Z$ and $h$ pairs suppressed by the mixing.
The decay width is given by
\begin{eqnarray}
\Gamma_H &=& \frac{M_H^3 \sin^2\alpha}{32 \pi v^2}
 \left[ 2 \left(1 -\frac{4 M_W^2}{M_H^2}+\frac{12 M_W^4}{M_H^4} \right)\sqrt{1-\frac{4M_W^2}{M_H^2}}
 + \left(1 -\frac{4 M_Z^2}{M_H^2}+\frac{12 M_Z^4}{M_H^4} \right)\sqrt{1-\frac{4M_Z^2}{M_H^2}} \right] \nonumber \\
 && + \frac{M_H^3 \cos^2\alpha}{32 \pi v_S^2}
 \left(1 -\frac{4 M_X^2}{M_H^2}+\frac{12 M_X^4}{M_H^4} \right)\sqrt{1-\frac{4M_X^2}{M_H^2}}  
 +\frac{g^2_{Hhh}}{8\pi M_H}\sqrt{1-\frac{4M^2_h}{M^2_H}}
 \nonumber \\
 && 
   + \Gamma(H \rightarrow  S_2  S_2^* \rightarrow S_2 l \nu).
\end{eqnarray}
Here, the three-body decay mode with one charged scalar $S_2$ being off-shell can be ignored when the lepton couplings to the charged scalar is small enough. We assume that this is the case, due to precision constraints associated with leptons as will be discussed in the later section.

We are interested in the case where 
 $X$ has a large annihilation cross section into two photons so that the gamma ray line at $135\gev$
 observed by Fermi LAT can be explained.
Taking the annihilation into a $W$ or $Z$ boson pair 
 through $h$ and $H$-exchange to be strongly suppressed due to a tiny mixing angle between SM Higgs boson and singlet scalars,
 namely, $|\sin\alpha| \ll 1$,
we obtain the desired thermal relic density and the necessary cross section into a photon pair
 for Fermi gamma-ray line near the resonance pole of the singlet-like scalar~\cite{resonance,resonance2}.
The necessary tiny mixing between the neutral scalars is given by $\lambda_4\sim\frac{1}{8\pi^2}\lambda_5\lambda_6\ln(M_{S^\pm_2}/\mu)$ from the one-loop corrections of the charged scalar $S_2$, provided that the tree-level mixing vanishes\footnote{One of the possibilities to realize a vanishing $\lambda_4$ at tree level is to put the model into extra dimensions. 
Namely, the charged scalar $S_2$ lives in the bulk of extra dimensions, while vector dark matter and $S_1$ are localized on a different location in extra dimensions than where Higgs doublet is localized.}.
For instance, for $\lambda_5\sim\lambda_6\sim 0.1$,  we can get $|\lambda_4|\sim 10^{-4}$, which is desirable for explaining both the relic density and Fermi gamma-ray line as will be shown later.
Therefore, near the resonance, $M_X \simeq M_H/2$, for the WIMP mass\footnote{
Of course, there is another resonance region $M_X \simeq M_h/2 \simeq 63$ GeV where
 the main annihilation mode is $XX \rightarrow b\bar{b}$.} around $135$ GeV,
 we need to take the singlet-scalar mass $M_H$ to be about $270$ GeV.

\begin{figure}[t]
\begin{center}
\includegraphics[scale=1.3]{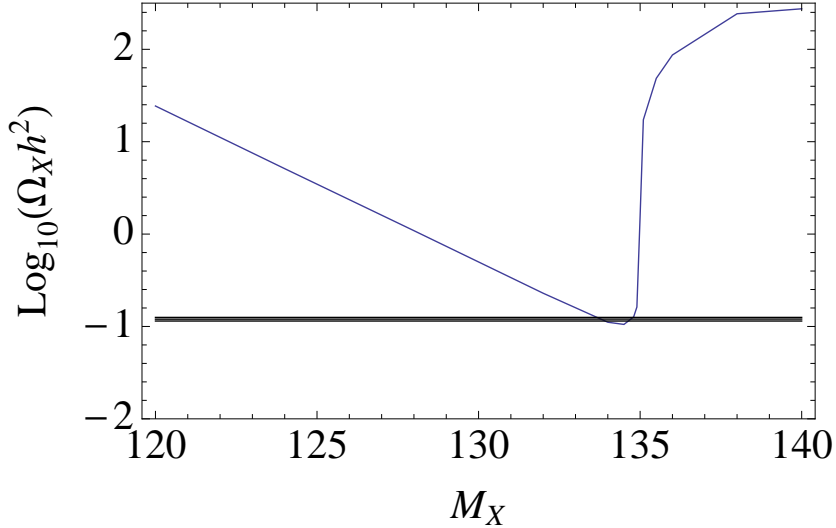}
\caption{The plot of relic density of the dark matter $\Omega h^2$ vs. its mass. 
Here we used $M_{S^\pm_2}=140\gev$, $v_S=1\tev$, $\sin\alpha=0.00047$ and $\lambda_6=0.2$. We can ignore the $\lambda_5$ dependence for the parameter region in our interest. The horizontal line is the relic density of cold dark matter revealed by the Planck result, $\Omega_{\rm CDM} h^2 = 0.1199\pm0.0027$~\cite{Ade:2013lta}.
}
\label{fig: omegah2}
\end{center}
\end{figure}

In Fig.~\ref{fig: omegah2}, we show that the right relic density of WIMP can be obtained 
by performing the thermal average of the annihilation cross section for all the dominant channels, with the procedure in Ref.~\cite{Gondolo:1990dk}.
We note that there is no bound on vector dark matter from direct detection in our model, due to a tiny Higgs mixing.

%%%%%%%%%%%%%%%%%%%%%%%
\subsection{Monochromatic photons from DM annihilation}
\label{subsec: line gamma}
%%%%%%%%%%%%%%%%%%%%%%%

Now we discuss the aspect of the indirect detection of vector WIMP. 
In addition to the $WW/ZZ$ annihilation channels,
 $X$ can also annihilate into two monochromatic photons with the effective interaction induced by the heavy charged scalar.
Hence, this diphoton mode can provide a source for the gamma-ray line observed by Fermi LAT.
When the singlet-like scalar $H$ has small couplings to $W$ and $Z$ bosons due to a tiny $\sin\alpha$ but it has a sizable coupling to the charged scalar  $S_2$,
the diphoton mode takes a larger branching fraction of the annihilation cross section than usually expected.

\begin{figure}[!t]
  \begin{center}
  \begin{tabular}{c}
   \includegraphics[width=0.7\textwidth]{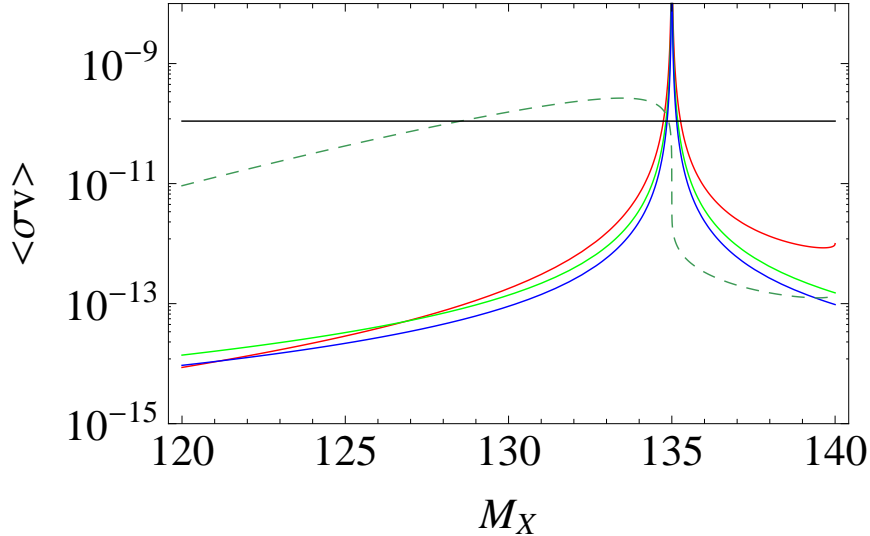}
     \end{tabular}
  \end{center}
  \caption{Thermal averaged annihilation cross section (Dashed line), given by the sum of $\gamma\gamma, WW, ZZ$ final states at the freeze-out temperature of dark matter, $T_f=M_X/20$. Solid lines are at  zero temperature (Red: $\gamma\gamma$, Blue: $WW$, Green: $ZZ$ final states). Here we used the same parameters in Fig.~\ref{fig: omegah2}.   The horizontal black line is the value of the DM annihilation cross section into two photons, $\VEV{\sigma v}_{\chi\chi \rightarrow \gamma\gamma} = 1.1\times 10^{-10}\gev^{-2}$,  needed for the Fermi-LAT gamma ray line.   
  }
  \label{crosssection}
\end{figure}

Keeping only the $H$-exchange contribution in Eq.~(\ref{sigmav:twogamma}) for a small Higgs mixing and using $s\simeq 4 M^2_X$,  we get an approximate form for the thermal averaged annihilation cross section into a photon pair at present as
\begin{eqnarray}
\langle\sigma v\rangle_{ \gamma\gamma}
 \simeq \frac{ \alpha_{\rm em}^2}{96 \pi^3 M_X^2} 
  \left|\frac{\lambda_6 \cos^2\alpha M_X^2 }{4 M_X^2-M_H^2+i M_H\Gamma_H }\right|^2
  \left| 1 - \frac{M_{S_2^+}^2}{M_X^2} f\left(\frac{M_{S_2^+}^2}{M_X^2}\right) \right|^2 .
\label{sigmav:twogamma:simplified}
\end{eqnarray}
In Fig.~\ref{crosssection}, we show, as a function of the DM mass, the annihilation cross sections into $WW,ZZ,\gamma\gamma$ at zero temperature, and the thermal averaged total annihilation cross section used to estimate $\Omega h^2$, respectively.
To explain the gamma ray line spectrum of the Fermi-LAT~\cite{Bringmann:2012vr,Weniger:2012tx},
for the Einasto dark matter profile, we require that
\dis{
\VEV{\sigma v}_{\chi\chi \rightarrow \gamma\gamma}&=(1.27\pm0.32^{+0.18}_{-0.28} )\times 10^{-27}\, {\rm cm}^3\, {\rm s}^{-1}, \\
&\approx1.1\times 10^{-10}\gev^{-2}.
}

In order to obtain the observed Fermi gamma-ray line together with the correct relic density,
 the small mixing angle, $|\sin\alpha| \ll 1$, is necessary. Furthermore, the annihilation into $WW$ and $ZZ$ modes have to be suppressed enough not to generate too many continuum photons~\cite{Buchmuller:2012rc,Cohen:2012me,Asano:2012zv}.
 In our case, the DM annihilation cross sections into $WW$ and $ZZ$ are suppressed at present for the parameters  which explain both the Fermi gamma-ray line and the relic density. For instance, for $M_X=134.74\,{\rm GeV}$, we obtain $\langle\sigma v\rangle_ {\gamma\gamma} = 1.09 \times 10^{-10}\gev^{-2}$ and 
$\Omega h^2=0.118$ while
$\langle\sigma v\rangle_{ZZ}/\langle\sigma v\rangle_ {\gamma \gamma}  =0.43$ and 
$\langle\sigma v\rangle_{WW}/\langle\sigma v\rangle_{ \gamma \gamma} = 0.28$. We note that the total DM annihilation cross section at present is smaller than thermal cross section, because the temperature effect shifts the peak of the resonance at freezeout towards a smaller DM mass as compared to the case with zero temperature \cite{Griest:1990kh}.

The same diagrams in Fig.~\ref{fig:XX2gg} applies to the annihilation of vector dark matter into $Z \gamma$ and
 loop induced $Z Z$ final states.  The annihilation $XX\rightarrow Z \gamma$ emits additional gamma line at $E_\gamma=114\gev$ and the resulting flux is suppressed by $0.21$ as compared to that of $130\gev$ gamma ray line~\cite{Choi:2012ap},
 while the loop induced annihilation into $ZZ$ is negligible.

%%%%%%%%%%%%%%%%%%%%%%%
\section{Higgs to diphoton rate}
\label{sec: Higgs diphoton}
%%%%%%%%%%%%%%%%%%%%%%%

The charged scalar $S_2$, introduced to explain the Fermi gamma-ray line, can give a positive or negative sizable contribution to the Higgs to diphoton rate\footnote{A similar discussion on the role of charged matter can be also found in the context of a scalar dark matter in Ref.~\cite{jijifan}.} due to the quartic coupling $\lambda_5$ to the SM Higgs field. 
In this section, we discuss the constraint on the modified decay rate of Higgs boson $h$ to diphotons from the recent measurements at the LHC.

We define the ratio of the Higgs production cross section times the branching fraction, $\mu_{\gamma\gamma}\equiv \frac{\sigma\times {\rm Br}_{\gamma\gamma}}{(\sigma\times {\rm Br}_{\gamma\gamma})_{SM}}$.
The reported signal strengths for the Higgs to diphoton rate from ATLAS and Multi-Variate-Analysis (MVA) of CMS data 
 are the following~\cite{March14},
\bea
\mu^{\rm ATLAS}_{\gamma\gamma} =1.65^{+0.34}_{-0.30}\,, \quad  \mu^{\rm CMS}_{\gamma\gamma}=0.78^{+0.28}_{-0.26}\,.
\eea
Following a similar method as in Refs.~\cite{Baglio:2012et,Espinosa:2012im} and assuming that the combined data is Gaussian, we have derived the combined value of the Higgs to diphoton rate as
\begin{equation}
\mu^{\rm combi}_{\gamma\gamma}=1.18\pm 0.20\,.
\end{equation}

\begin{figure}[t]
\begin{center}
\includegraphics[scale=1.0]{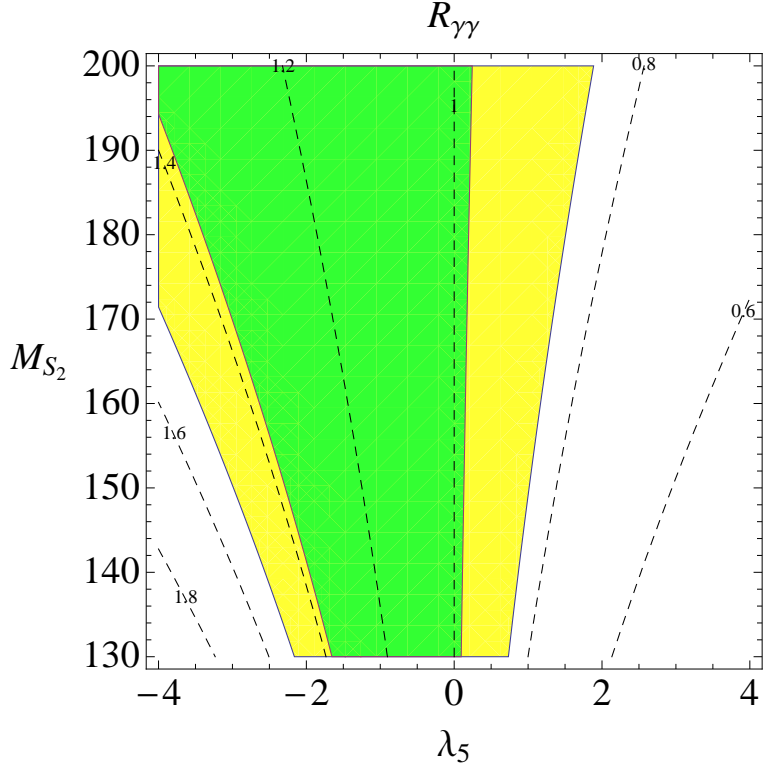}
\caption{
Contours of Higgs to diphoton rate on $\lambda_5 - M_{S_2^\pm}$ plane.
The green and yellow regions correspond to the combined LHC bounds on $R_{\gamma\gamma}$ at $68\%$ and $90\%$ C. L., respectively.
}
\label{fig:Higgsdiphoton}
\end{center}
\end{figure}

The SM-like Higgs boson decay width $\Gamma(h\rightarrow \gamma \gamma$) is given by~\cite{HHG}
\be
\Gamma(h\rightarrow \gamma \gamma) = \frac{G_F \alpha^2 M^3_h}{128\sqrt{2}\pi^3}
\left| g_W A_1 (\tau_W)+ g_t Q^2_t N_c A_{1/2}(\tau_t) + g_h A_0 (\tau_{S_2^{\pm}})\right|^2,
\ee
with loop functions
\bea
A_1(x)&=& -x^2 [2x^{-2}+3x^{-1}+3(2x^{-1}-1)f(x^{-1}) ], \\
A_{1/2}(x)&=& 2x^2 [x^{-1} +(x^{-1}-1) f(x^{-1}) ], \\
A_0(x)&=& -x^2 [x^{-1}-f(x^{-1})], \\
f(x)&=&\bigg\{ \begin{array}{ll}\arcsin^2\sqrt{x}\quad\quad {\rm for} \,\, x\leq 1 \\ 
-\frac{1}{4}(\ln((1+\sqrt{1-x^{-1}})/(1+\sqrt{1-x^{-1}}))-i \pi)^2\quad {\rm for}\,\,   x>1,
\end{array}
\eea
and $\tau_i=4M^2_i/M_h^2$. $Q_t=\frac{2}{3}$, $N_c=3$ for top quark. 
$g_W$ and $g_t$ are the Higgs trilinear couplings to $W$ gauge boson and 
 top quark normalized to the ones of the SM, and in our case those are almost $1$.
The Higgs coupling to the charged scalar boson is 
\begin{equation}
 g_h=\frac{M_W}{g \,M^2_{S_2^{\pm}}} \,v \lambda_5.
\end{equation}
By taking the ratio to the SM value, we obtain 
\begin{equation}
R_{\gamma\gamma}\equiv\frac{\Gamma(h\rightarrow \gamma \gamma)}{\Gamma(h\rightarrow \gamma \gamma)_{\rm SM}}
 =
\left| 1 + \frac{g_h A_0 (\tau_{S_2^{\pm}})}{g_W A_1 (\tau_W)+ g_t Q^2_t N_c A_{1/2}(\tau_t)}\right|^2,
\end{equation}
which is a function of $\lambda_5$ and $M_{S_2^\pm}$.
When the production cross section of $h$ is SM Higgs-like, the above ratio is approximated to the signal strength  measured at the LHC, that is, $R_{\gamma\gamma}\simeq\mu_{\gamma\gamma}$.

In Fig.~\ref{fig:Higgsdiphoton}, we depict the parameter space $(\lambda_5,M_{S^\pm_2})$, showing the contours of the $h$ decay to diphoton rate with other Higgs couplings being assumed the same as in the SM. We find that, from the combined value of ATLAS and CMS (MVA) diphoton signal strengths at $90\%$ C.L., the extra quartic coupling is constrained to $-2.5\lesssim\lambda_5\lesssim 0.7$ for $M_{S^\pm_2}=140\,{\rm GeV}$. The heavier the charged scalar,  the larger values of the extra quartic coupling $\lambda_5$ are allowed.

%%%%%%%%%%%%%%%%%%%%%%%
\section{Constraints on charged scalar}
\label{sec: constraints on charged scalar}
%%%%%%%%%%%%%%%%%%%%%%%

We have introduced the lepton couplings to the charged scalar $S_2$ by gauge invariant terms, $f_{ij}L_i^c \cdot L_j S_2$, so $S_2$ could be unstable.
In this section, we discuss the phenomenology of the charged scalar from the indirect limits
 and the collider search for charged particles at the LHC. 

The lepton couplings to the charged scalar are similar to lepton number (R-parity) violating terms
 in the minimal supersymmetric standard model, $\Delta W= \lambda_{ijk} L_i\cdot L_j E_k^c$,
 so the same bounds from precision measurements are applied to them.
The bounds from the charged current universality are  $|f_{12}|<0.04(M_{S^\pm_2}/(100\,{\rm GeV}))$, and the constraints from $R_\tau=\Gamma(\tau\rightarrow e \nu {\bar \nu})/\Gamma(\tau\rightarrow \mu \nu{\bar \nu})$ or $R_{\tau\mu}=\Gamma(\tau\rightarrow \mu \nu {\bar\nu})/\Gamma(\mu\rightarrow e\nu {\bar\nu})$ are $|f_{ij}|<0.05(M_{S^\pm_2}/(100\,{\rm GeV}))$, and $\nu_\mu$ deep inelastic scattering gives the bound, $|f_{12}|<0.02(M_{S^\pm_2}/(100\,{\rm GeV}))$~\cite{RPV}.
Other lepton flavor violation $Br(\mu \rightarrow e \gamma)$ also gives a similar bounds~\cite{Dicus:2001ph}.
The charged scalar couplings contribute to the effective tree-level Fermi coupling in $\mu$-decay by 
$G_\mu/\sqrt{2}=g^2/(8M^2_W)+|f_{12}|^2/(8M^2_{S^\pm_2})$, but it gives a less stringent limit than the bounds quoted above \cite{barger}.
The lepton Yukawa couplings gives a negative contribution\footnote{There was an error in the previous works on the $LLS_2$
 coupling~\cite{Dicus:2001ph} which showed a positive contribution to the muon anomalous magnetic moment.} 
 to the muon anomalous magnetic moment as ~\cite{Kim:2001se,Nebot:2007bc} 
\be
\Delta a_\mu =-\frac{m^2_\mu}{96\pi^2} \frac{1}{M^2_{S^\pm_2}} (|f_{12}|^2+|f_{23}|^2).
\ee
Then, using the bounds from precision measurements, we get a very small contribution, $|\Delta a_\mu|<3.45\times 10^{-12}$. We note that as far as the electroweak precision bounds on $f_{ij}$ are satisfied for $M_{S^\pm_2}\gtrsim 130\,{\rm GeV}$, the continuum photons coming from the three-body DM annihilation into $ S_2 l\nu$ can be suppressed enough.

New particles with electroweak charges have been searched for at colliders.
The stringent bounds on the charged scalar come from the direct slepton pair production
 at the LHC~\cite{ATLASslepton,CMSslepton}, where a left-handed slepton can decay into lepton and neutralino. 
The opposite-sign dilepton search with the same-flavor channel at CMS excludes slepton masses
 between $110\,{\rm GeV}$ and $275\,{\rm GeV}$ for massless neutralino~\cite{CMSslepton}. 
But, in our case, the charged scalar can decay into all the charged leptons: $S^-_2\rightarrow e{\bar\nu}_{\mu,\tau}, \mu {\bar\nu}_{e,\tau}$, and $\tau {\bar\nu}_{e,\mu}$. 
Since the lepton coupling matrix, $f_{ij}$, is anti-symmetric, at least two different flavors always appear in the decay product of the charged scalar. Therefore, the CMS mass limit with the same-flavor channel scales down or does not apply, depending on the branching fraction of the same-flavor decay mode. Currently, the most stringent constraint on the charged scalar mass comes from the LEP exclusion limit 
up to $95\,{\rm GeV}$~\cite{LEPslepton}.

%%%%%%%%%%%%%%%%%%%%%%%
\section{Vacuum stability}
\label{sec: vacuum stability}
%%%%%%%%%%%%%%%%%%%%%%%

The discovered scalar boson with $126 \,{\rm GeV}$ mass has been shown to have very similar properties to the SM Higgs boson with more precision \cite{Higgsfit}. Although we need more data to confirm the properties of the Higgs boson, we assume that the discovered scalar boson is SM Higgs-like.  In this case, a small Higgs quartic coupling leads to a problem of vacuum instability below the Planck scale \cite{VSB}, requiring new physics beyond the SM \footnote{We note that when the top pole mass is smaller than $171\,{\rm GeV}$, the electroweak vacuum could be absolutely stable until the Planck scale without new physics \cite{VSB,giudice}.}. In this section, we discuss the effect of the additional quartic couplings between the Higgs boson and extra scalars in the model, taking account of dark matter constraints from Fermi gamma-ray line, Higgs boson data and other collider bounds, discussed in the previous sections.

\begin{figure}[!tb]
  \begin{center}
   \includegraphics[width=6cm]{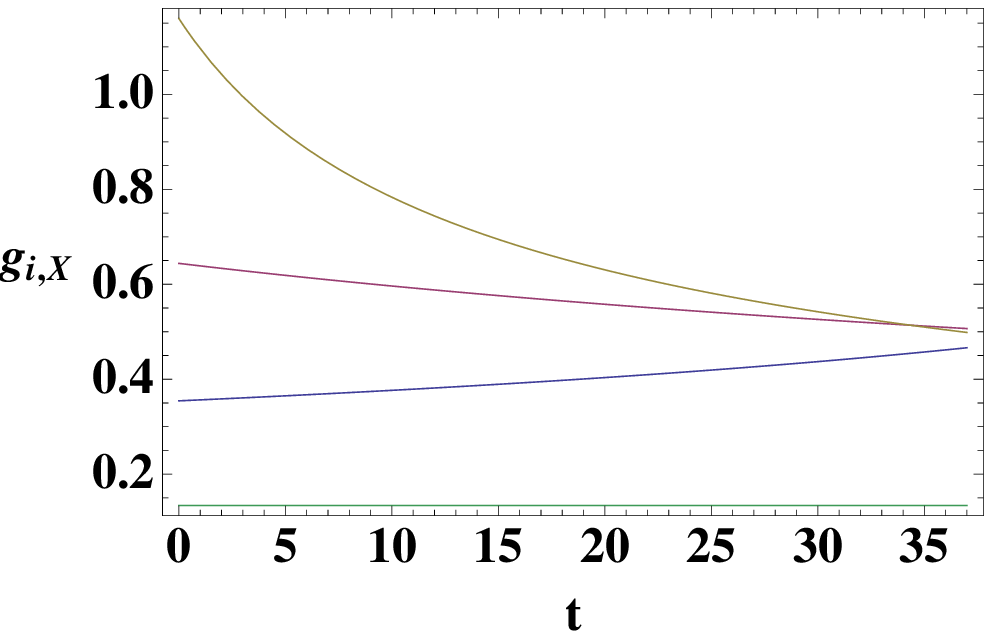}
   \includegraphics[width=6cm]{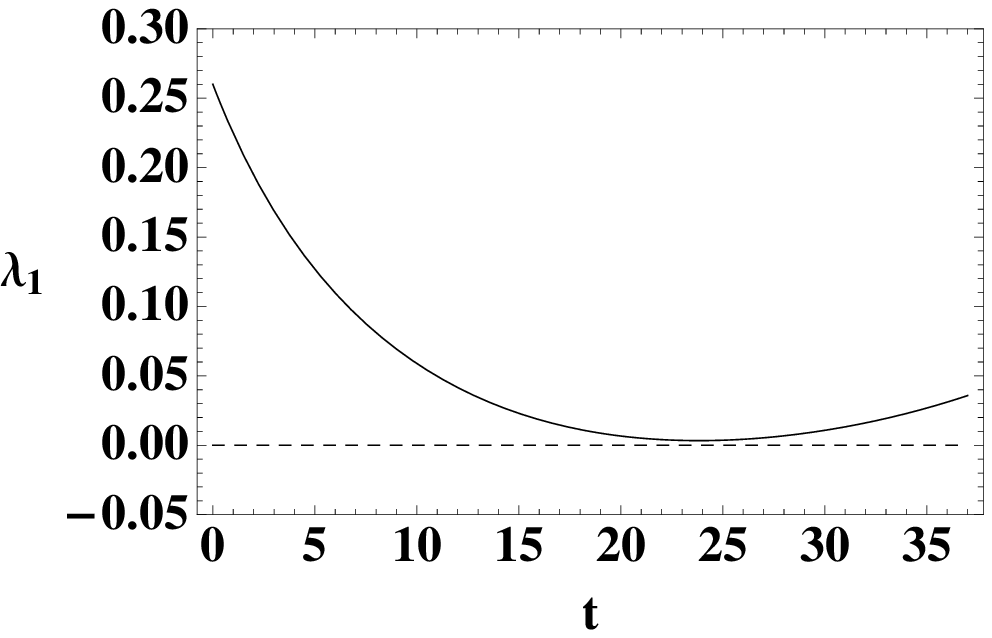} \\
   \includegraphics[width=6cm]{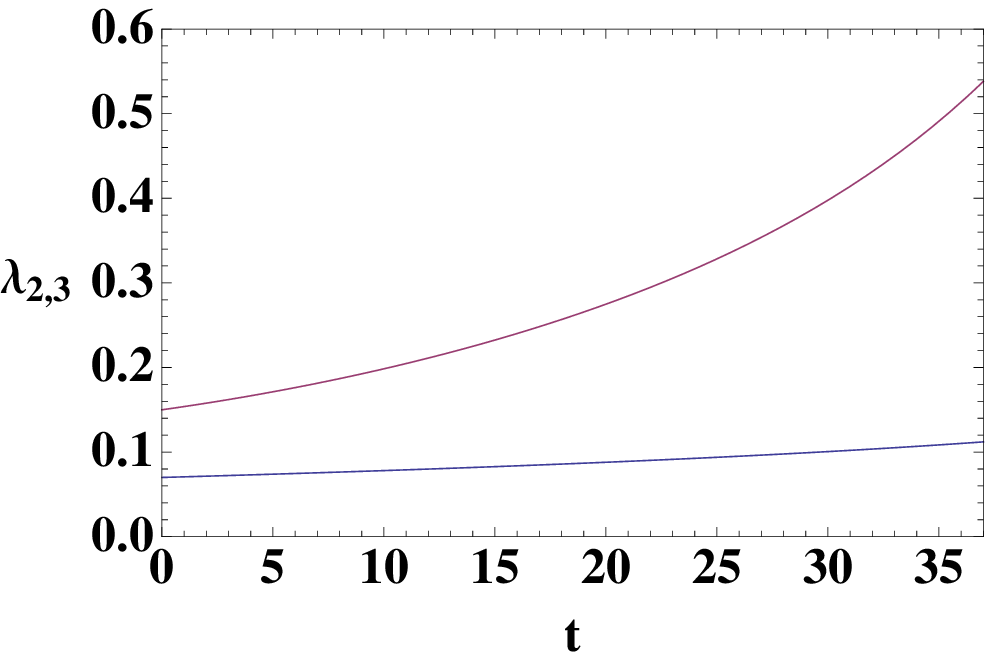}
   \includegraphics[width=6cm]{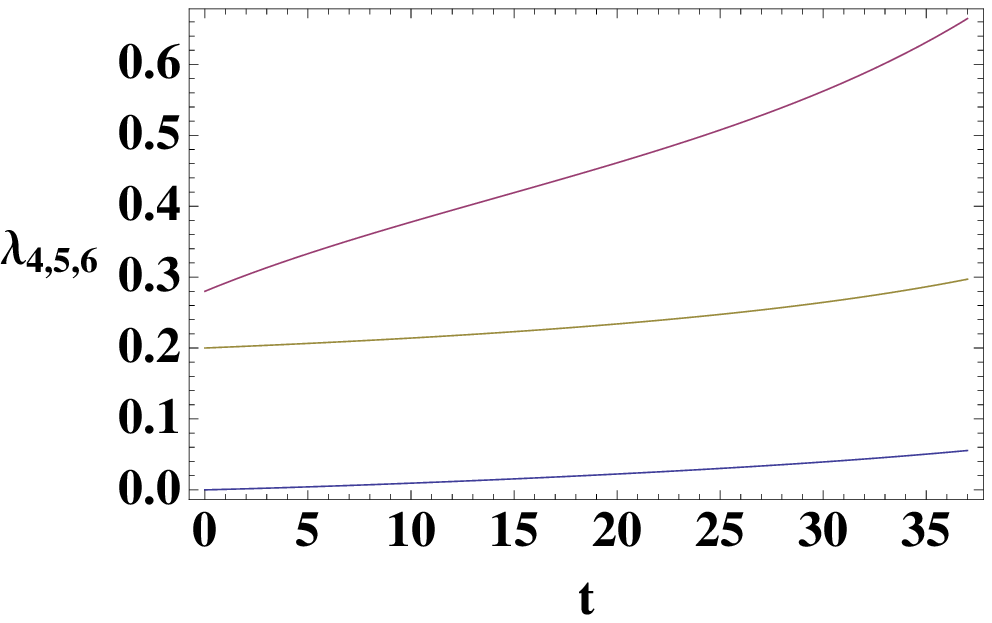}
 \end{center}
  \caption{Running couplings as a function of $t=\ln(\mu/M_t)$. 
  We have chosen $\lambda_1=0.26, \lambda_2=0.07, \lambda_3=0.15, \lambda_4=-0.0001, \lambda_5=0.28, \lambda_6=0.2$
 and $g_X=0.134$ at the top pole mass, $M_t=173\,{\rm GeV}$, which leads to $M_X=134\,{\rm GeV}$, $M_H=270\,{\rm GeV}$, $M_h=126\,{\rm GeV}$ and $\sin\alpha=0.0005$ with $v_S=1\,{\rm TeV}$. The lepton couplings to the charged scalar are ignored in the RG analysis.  }
  \label{fig:run}
\end{figure}

In our model, 
 it is possible to have a sizable shift in the Higgs quartic coupling in the presence of
 the mixing with a singlet scalar~\cite{giudice,lebedev,batell} as follows,
\be
\lambda_{\rm eff}=\lambda_1-\delta\lambda
\ee
with
\be
\delta\lambda = \frac{(M^2_H-M^2_h)^2\sin^2\alpha\cos^2\alpha}{v^2(M^2_H\cos^2\alpha +M^2_h \sin^2\alpha)}.
\ee
In the decoupling limit of a heavy singlet-like scalar with $M_H\gg M_h$,
 the tree-level shift is approximated to $\delta\lambda\simeq M^2_H\sin^2\alpha/(v^2)\simeq \lambda^2_4/\lambda_2$~\cite{giudice,lebedev}.
In this paper, however, we take the singlet-like scalar mass to be close to the resonance pole, $M_H\sim 2 M_X\sim 270\,{\rm GeV}$. 
Furthermore, since $|\sin\alpha|\ll 1$ for the correct relic density at the resonance, the tree-level shift in our case becomes $\delta\lambda\simeq 0.7 \sin^2\alpha$, which is extremely small.

\begin{figure}[!t]
  \begin{center}
  \begin{tabular}{c}\includegraphics[width=6cm]{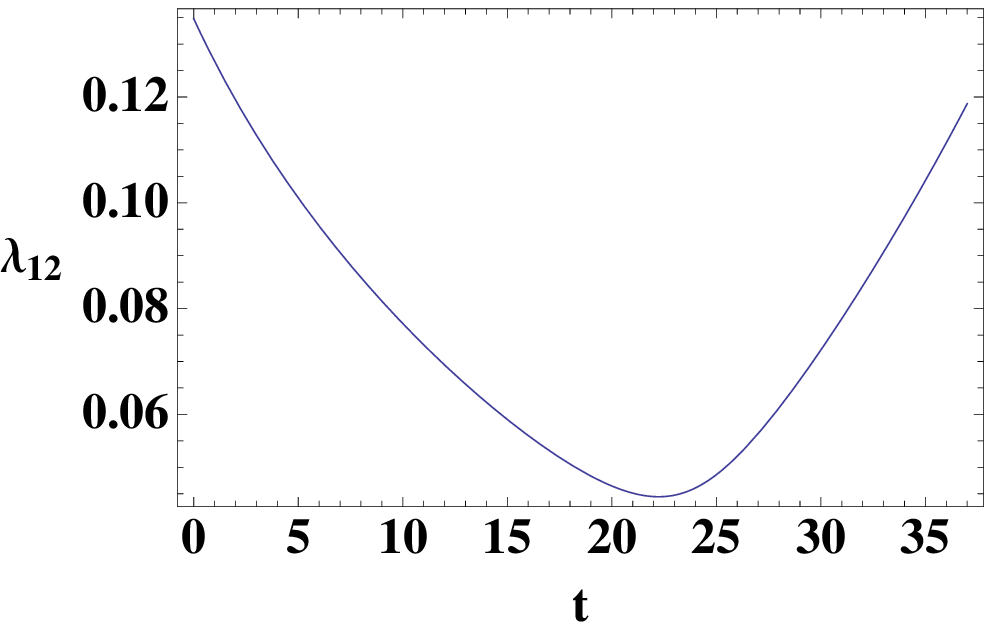}
   \includegraphics[width=6cm]{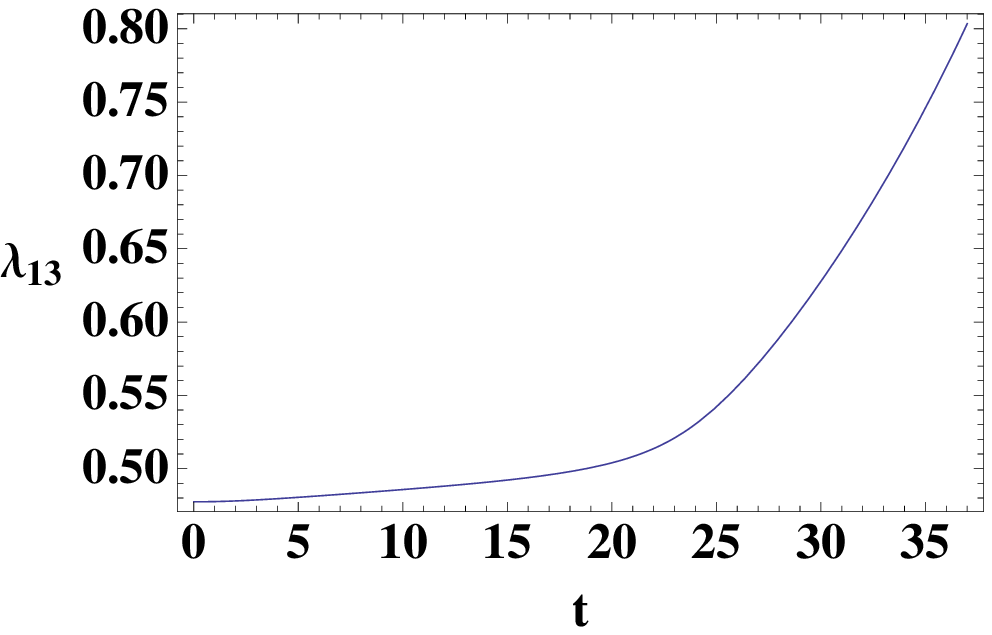} \\
   \includegraphics[width=6cm]{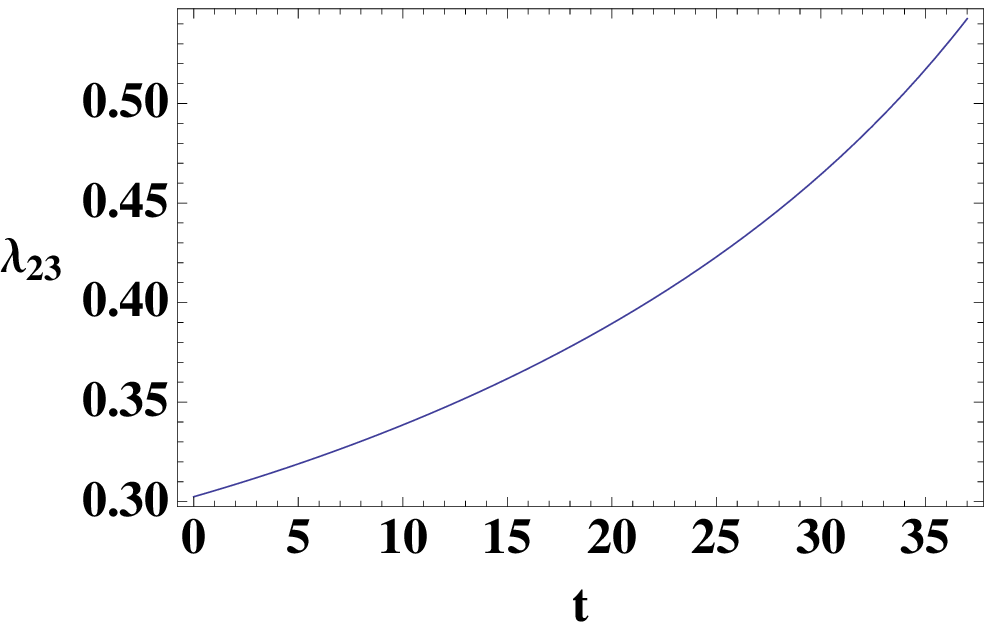}
   \includegraphics[width=6cm]{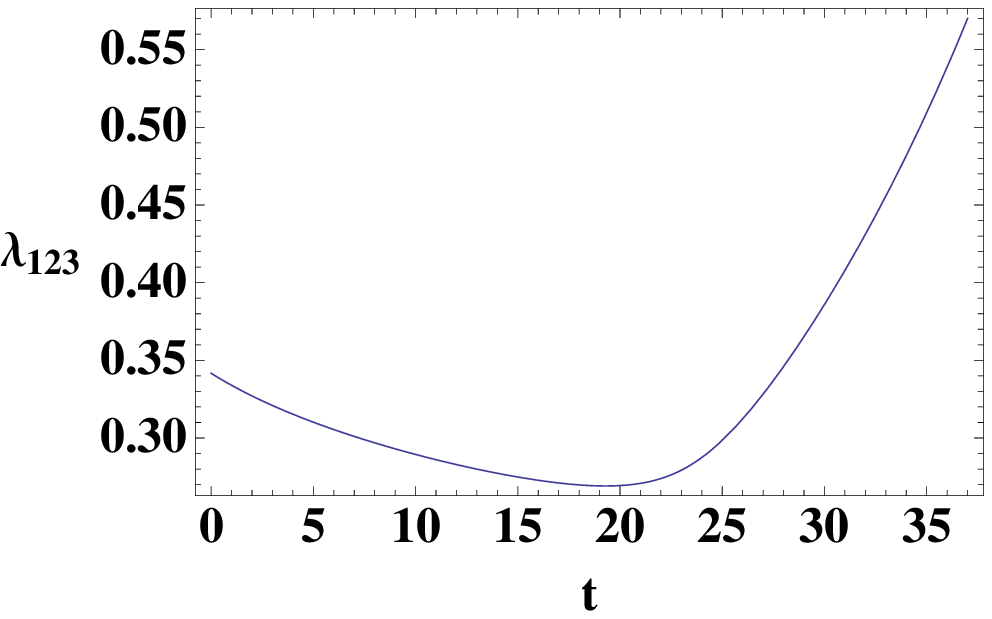}
   \end{tabular}
  \end{center}
  \caption{Vacuum stability conditions: $\lambda_{12}>0, \lambda_{13}>0, \lambda_{23}>0$ and $\lambda_{123}>0$.
   The same parameters are chosen as in Fig.~\ref{fig:run}. }
  \label{fig:vacuumstability}
\end{figure}

Now we consider the RG effect on the Higgs quartic coupling.
As shown in Appendix~\ref{sec: rge}, 
 there are positive contributions to the beta function of the Higgs quartic coupling, $\lambda_4$ and $\lambda_5$, in the RG equations, so the vacuum stability can be improved as compared to the SM. 
But, from the results of the previous sections, the quartic coupling $\lambda_4$ between the Higgs doublet and the singlet scalar must be small because of the relic density condition, hence its contribution to the running of the Higgs quartic coupling is negligible. 
On the other hand, a sizable quartic coupling $\lambda_5$ between the Higgs doublet and the charged scalar is allowed,
 being consistent with the observed Higgs boson decay rate to diphoton. 
 
If $\lambda_5$ is positive, it can help increase the vacuum instability scale, without violating the new vacuum stability conditions of extra scalars. We note, however, that if $\lambda_5$ is negative and large as suggested by Higgs data, it could increase the Higgs quartic coupling by the RG further, but perturbativity bound and extra vacuum stability conditions strongly restrict this possibility.  In Fig.~\ref{fig:run}, we show the running couplings until the Planck scale for the low-energy couplings including a positive $\lambda_5$, that are consistent with Fermi gamma-ray line, relic density, Higgs diphoton data and indirect and collider bounds. In Fig.~\ref{fig:vacuumstability}, the vacuum stability conditions are shown to be satisfied until the Planck scale, for the same parameters as in Fig.~\ref{fig:run}.

%%%%%%%%%%%%%%%%%%%%%%%
\section{Conclusion}
\label{sec: conclusion}
%%%%%%%%%%%%%%%%%%%%%%%

We have proposed a renormalizable model of vector dark matter
 where the extra $U(1)_X$ gauge boson is a dark matter candidate and it interacts with the SM particles through the Higgs portal term, namely,
 the mixing between the $U(1)_X$ breaking singlet scalar and the SM Higgs doublet. 
If the Higgs mixing is small enough,
the DM annihilations into $W$ and $Z$ boson pairs at the resonance pole of the singlet-like scalar can reproduce the correct thermal relic density without overproducing continuum photons.
In the presence of a quartic coupling between the singlet scalar and the charged scalar $S_2$, vector dark matter also annihilates into a photon pair with a sizable branching fraction at the same singlet resonance. 

 As long as the couplings of the charged scalar to the SM leptons are small enough, i.e. $f_{ij} < {\cal O} (10^{-2})$, we showed that  all the electroweak precision constraints concerning leptons are satisfied. 
 Even though it would be very difficult to find the singlet scalar at colliders due to a tiny mixing with the Higgs, the charged scalar would be accessible at the LHC or linear colliders, due to a distinct signature that two opposite-sign leptons of different flavors are equally produced from the charged scalar decay.
We have also shown that the vacuum stability bounds are satisfied until the Planck scale, due to a sizable Higgs coupling to the charged scalar, which is allowed by the current Higgs diphoton data.

\section*{Acknowledgments}

K.-Y.C was supported by Basic Science Research Program through the National Research Foundation of Korea (NRF) funded by the Ministry of Education, Science and Technology (No. 2011-0011083).
K.-Y.C acknowledges the Max Planck Society (MPG), the Korea Ministry of
Education, Science and Technology (MEST), Gyeongsangbuk-Do and Pohang
City for the support of the Independent Junior Research Group at the Asia Pacific
Center for Theoretical Physics (APCTP).
This work of O.S. is in part supported by scientific research grants from Hokkai-Gakuen. 
O.S. thank Physik-Department T30d, Technische Universitat Munchen 
 for their warm hospitality during his visit supported
 by Japan Society for the Promotion of Science and Deutscher Akademischer Austauschdienst, 
 where this work has been completed.

%======================================%
%<<<<<<<<<<< appendix >>>>>>>>>>>>>%
%======================================%
\appendix
%%%%%%%%%%%%%%%%%%%%%%%
\section{Scalar interaction vertices}
\label{sec: vertex}
%%%%%%%%%%%%%%%%%%%%%%%

Gauge interactions:
\begin{eqnarray}
{\cal L}_{int} &=& g_X^2  X^2 v_S (H \cos\alpha - h \sin\alpha )  \nonumber \\
   && + \frac{g_2^2}{4 c_W^2} (2 c_W^2 W^- W^+ +  Z^2) v (h \cos\alpha + H \sin\alpha ).
\end{eqnarray}

Scalar interactions:
\dis{
-{\cal L}_{int} =& \frac{1}{2}  (v \lambda_1 \cos^3\alpha+v \lambda_4 \sin^2\alpha \cos\alpha
 -v_S \lambda_2 \sin^3\alpha - v_S \lambda_4 \sin\alpha \cos^2\alpha ) h^3 \\
& +\frac{1}{8} (v \sin\alpha (3 \lambda_1+\lambda_4 )
 +3 v \sin3\alpha (\lambda_1-\lambda_4 )+v_S \cos\alpha (3 \lambda_2+\lambda_4 )
 +3 v_S \cos3\alpha (\lambda_4-\lambda_2) ) h^2 H  \\
 &+\frac{1}{8} (v \cos\alpha (3 \lambda_1+\lambda_4)-3 v \cos3\alpha (\lambda_1-\lambda_4)-v_S \sin\alpha (3 \lambda_2+\lambda_4 )+3 v_S \sin3\alpha (\lambda_4-\lambda_2 )   ) h H^2  \\
 &+ \frac{1}{2}  (v \lambda_1 \sin^3\alpha+v \lambda_4 \sin\alpha \cos^2\alpha
 +v_S \lambda_2 \cos^3\alpha + v_S \lambda_4 \cos\alpha \sin^2\alpha ) H^3\\
& + ( v \lambda_5 \cos\alpha - v_S \lambda_6 \sin\alpha ) h S^+ S^-  
+ ( v \lambda_5 \sin\alpha + v_S \lambda_6 \cos\alpha ) H S^+ S^- \\
&+\frac18(\cos^4\alpha \lambda_1+\sin^4\alpha \lambda_2+2\cos^2\alpha\sin^2\alpha \lambda_4 )h^4+\frac18(\sin^4\alpha \lambda_1+\cos^4\alpha \lambda_2+2\cos^2\alpha\sin^2\alpha \lambda_4 )H^4\\
&+ \frac12(\cos^3\alpha \sin\alpha (\lambda_1 - \lambda_4) - \cos\alpha\sin^3\alpha(\lambda_2-\lambda_4)  )h^3H \\
&+ \frac12(\cos\alpha \sin^3\alpha (\lambda_1 - \lambda_4) - \cos^3\alpha\sin\alpha(\lambda_2-\lambda_4)  )hH^3 \\
&+\frac14( (\cos^4\alpha-4\cos^2\alpha\sin^2\alpha+\sin^4\alpha  )  \lambda_4 +3\cos^2\alpha\sin^2\alpha (\lambda_1+\lambda_2) )  h^2H^2\\
&+\frac12 (\cos^2\alpha  \lambda_5 +\sin^2\alpha \lambda_6 )h^2  S^+ S^- +\frac12 (\sin^2\alpha  \lambda_5 +\cos^2\alpha \lambda_6 )H^2  S^+ S^- \\
& + \sin\alpha \cos\alpha(\lambda_5-\lambda_6) h H  S^+ S^- +\frac12\lambda_3 |S^+S^-|^2.
}
%

%%%%%%%%%%%%%%%%%%%%%%%
\section{Renormalization group equations}
\label{sec: rge}
%%%%%%%%%%%%%%%%%%%%%%%

The running of the coupling $p_i$ is governed by the RG equation, $\frac{\partial p_i}{\partial t}=\beta_{p_i}$ with $\beta_{p_i}$ being the corresponding beta function and $t\equiv \ln(\mu/m_t)$. 
The beta functions for scalar quartic couplings with $\kappa\equiv 16\pi^2$ are
\bea
\kappa\beta_{\lambda_1}&=&12 \lambda^2_1 +(12y^2_t - 9 g^2-3 g^{\prime 2})\lambda_1 -12 y^4_t +\frac{9}{4} g^4+\frac{3}{4} g^{\prime 4}+\frac{3}{2}g^2 g^{\prime 2} +2 \lambda^2_4 +2\lambda^2_5, \\
\kappa\beta_{\lambda_2}&=& 10 \lambda^2_2 + 4\lambda^2_4 +2\lambda^2_6-12 g^2_X \lambda_2+12 g^4_X, \\
\kappa\beta_{\lambda_3}&=& 10 \lambda^2_3+4 \lambda^2_5 +2 \lambda^2_6+\Big(4 {\rm Tr}(f^\dagger f)-12 g^{\prime 2}\Big)\lambda_3+12 g^{\prime 4}-4 {\rm Tr}(f f^\dagger f f^\dagger), \\
\kappa\beta_{\lambda_4}&=&(6\lambda_1+4\lambda_2+4\lambda_4)\lambda_4 +2\lambda_5 \lambda_6 +\Big(6 y^2_t-\frac{3}{2}g^{\prime 2}-\frac{9}{2}g^2-6 g^2_X\Big)\lambda_4, \\
\kappa\beta_{\lambda_5}&=& (6\lambda_1+4\lambda_3+4\lambda_5)\lambda_5 +2\lambda_4\lambda_6 +\Big(6y^2_t+ 2{\rm Tr}(f^\dagger f)-\frac{15}{2}g^{\prime 2}-\frac{9}{2} g^2\Big)\lambda_5 + 3 g^{\prime 4}, \\
\kappa\beta_{\lambda_6}&=& 4 (\lambda_2+\lambda_3+\lambda_6)\lambda_6 + 4\lambda_4 \lambda_5 +\Big(2{\rm Tr}(f^\dagger f)-6g^{\prime 2}- 6g^2_X\Big)\lambda_6.
\eea
The beta functions for the top Yukawa coupling and the lepton Yukawa couplings to the charged scalar are
\bea
\kappa \beta_{y_t}&=& y_t\Big(\frac{9}{2}y^2_t- 8g^2_3 -\frac{9}{4}g^2-\frac{17}{12} g^{\prime 2}\Big), \\
\kappa \beta_{f_{ij}}&=& 4(f f^\dagger f)_{ij} + f_{ij} \Big(4{\rm Tr}(f^\dagger f)-\frac{9}{2} g^2 - \frac{3}{2} g^{\prime 2}\Big).
\eea
Here, we have ignored the charged lepton Yukawa couplings to the SM Higgs field.
When a single lepton coupling to the charged scalar, e.g. $f\equiv |f_{12}|$,  is dominant, the RG equation for that becomes
\be
\kappa \beta_f= f\Big(12 f^2-\frac{9}{2}g^2-\frac{3}{2}g^{\prime 2}\Big).
\ee

The beta functions for the gauge couplings are
\bea
\kappa\beta_{g'} = \frac{43}{6}g^{\prime 3},\quad \kappa\beta_g=-\frac{19}{6} g^3,\quad \kappa\beta_{g_3}=-7 g^3_3,\quad \kappa\beta_{g_X}=\frac{1}{3}g^3_X.
\eea

%
%======================================%
%<<<<<<<<<<< bibliography >>>>>>>>>>>>>%
%======================================%

%%%%%%%%%%%%%%%%%%%%%%%%%%%%%%%%%%%%%%%%%%%%%%%%%%%%%%%%%%%%


\begin{thebibliography}{99}
%%%%%%%%%%%%%%%%%%%%%%%%%%%%%%%%%%%%%%%%%%%%%%%%%%%%%%%%%%%%


\bibitem{DMreview}
G.~Jungman, M.~Kamionkowski and K.~Griest, Phys.\ Rept.\ {\bf 267}, 195 (1996);\\
C.~Munoz, Int.\ J.\ Mod.\ Phys.\ A {\bf 19}, 3093 (2004).

%%%%%%%%%%%%%%%  Fermi 130  %%%%%%%%%%%%%%%%%%%%
%\cite{Bringmann:2012vr}
\bibitem{Bringmann:2012vr}
  T.~Bringmann, X.~Huang, A.~Ibarra, S.~Vogl and C.~Weniger,
  %``Fermi LAT Search for Internal Bremsstrahlung Signatures from Dark Matter Annihilation,''
  JCAP {\bf 1207}, 054 (2012).
  %\cite{Weniger:2012tx}
\bibitem{Weniger:2012tx}
  C.~Weniger,
  %``A Tentative Gamma-Ray Line from Dark Matter Annihilation at the Fermi Large Area Telescope,''
  JCAP {\bf 1208}, 007 (2012).
%
\bibitem{Tempel}  
  E.~Tempel, A.~Hektor and M.~Raidal,
  %``Fermi 130 GeV gamma-ray excess and dark matter annihilation 
  %in sub-haloes and in the Galactic centre,''
  JCAP {\bf 1209}, 032 (2012) [Addendum-ibid.\  {\bf 1211}, A01 (2012)].
\bibitem{Su:2012ft}
  M.~Su and D.~P.~Finkbeiner,
  %``Strong Evidence for Gamma-ray Line Emission from the Inner Galaxy,''
 arXiv:1206.1616 [astro-ph.HE].
%
%%%%%%%%  Analysis by Fermi coll. %%%%%%%%%%%%%%%%
%\cite{Bloom:2013mwa}
\bibitem{Bloom:2013mwa} 
  E.~Bloom {\it et al.}  [On Behalf of the Fermi-LAT Collaboration],
  %``Search of the Earth Limb Fermi Data and Non-Galactic Center
  % Region Fermi Data for Signs of Narrow Lines,''
  arXiv:1303.2733 [astro-ph.HE].
%
%%%%%%%%  Pulser  %%%%%%%%%%%%%%%%%
%\cite{Aharonian:2012cs}
\bibitem{Aharonian:2012cs}
  F.~Aharonian, D.~Khangulyan and D.~Malyshev,
  %``Cold ultrarelativistic pulsar winds as potential sources of 
  %galactic gamma-ray lines above 100 GeV,''
  arXiv:1207.0458 [astro-ph.HE].
%
%%%%%%%%  Fermi bubble  %%%%%%%%%%%%%%%%%
\bibitem{Su:2010qj}
  M.~Su, T.~R.~Slatyer and D.~P.~Finkbeiner,
  %``Giant Gamma-ray Bubbles from Fermi-LAT: AGN Activity or Bipolar Galactic Wind?,''
  Astrophys.\ J.\  {\bf 724}, 1044 (2010).
\bibitem{Profumo:2012tr}
  S.~Profumo and T.~Linden,
  %``Gamma-ray Lines in the Fermi Data: is it a Bubble?,''
  JCAP {\bf 1207}, 011 (2012).
%
%%%%%%%%  Instrumental %%%%%%%%%%%%%%%%%%%%%%%%%%
%\cite{Whiteson:2012hr}
\bibitem{Whiteson:2012hr}
  D.~Whiteson,
  %``Disentangling Instrumental Features of the 130 GeV Fermi Line,''
  JCAP {\bf 1211}, 008 (2012).
%  [arXiv:1208.3677 [astro-ph.HE]].
\bibitem{Finkbeiner:2012ez}
  D.~P.~Finkbeiner, M.~Su and C.~Weniger,
  %``Is the 130 GeV Line Real? A Search for Systematics in the Fermi-LAT Data,''
  JCAP {\bf 1301}, 029 (2013).
%  [arXiv:1209.4562 [astro-ph.HE]].
\bibitem{Whiteson:2013cs}
  D.~Whiteson,
  %``Searching for Spurious Solar and Sky Lines in the Fermi-LAT Spectrum,''
  arXiv:1302.0427 [astro-ph.HE].
%
%%%%%%%%  Earth Limb   %%%%%%%%%%%%%%%%%%%%%%%%%%
%\cite{Hektor:2012ev}
\bibitem{Hektor:2012ev}
  A.~Hektor, M.~Raidal and E.~Tempel,
  %``Fermi-LAT gamma-ray signal from Earth Limb, systematic detector
  % effects and their implications for the 130 GeV gamma-ray excess,''
  arXiv:1209.4548 [astro-ph.HE]. 


%%%%%%%%  Analysis by Fermi coll. %%%%%%%%%%%%%%%%
\bibitem{Ackermann:2012qk} 
  M.~Ackermann {\it et al.}  [LAT Collaboration],
  %``Fermi LAT Search for Dark Matter in Gamma-ray Lines and the Inclusive Photon Spectrum,''
  Phys.\ Rev.\ D {\bf 86}, 022002 (2012).
%  [arXiv:1205.2739 [astro-ph.HE]].

%%%%%%%%  None from HESS        %%%%%%%%%%%%%%%%%
%\cite{Abramowski:2013ax}
\bibitem{Abramowski:2013ax} 
  A.~Abramowski {\it et al.}  [H.E.S.S. Collaboration],
  %``Search for photon line-like signatures from Dark Matter annihilations with H.E.S.S,''
  Phys.\ Rev.\ Lett.\  {\bf 110}, 041301 (2013).

%%%  Models for Fermi 130 %%%%%%%%%%%%%%%%%%%%%
%%%%%%%%% Large coupling for Fermi gamma-ray line %%%%%%%%%
%\cite{Cline}
\bibitem{Cline} 
  J.~M.~Cline,
  %``130 GeV dark matter and the Fermi gamma-ray line,''
  Phys.\ Rev.\ D {\bf 86}, 015016 (2012).
%\cite{Choi:2012ap}
\bibitem{Choi:2012ap}
  K.~-Y.~Choi and O.~Seto,
  %``A Dirac right-handed sneutrino dark matter and its signature in the gamma-ray lines,''
  Phys.\ Rev.\ D {\bf 86}, 043515 (2012)
   [Erratum-ibid.\ D {\bf 86}, 089904 (2012)].

   
   
%\cite{Tulin:2012uq}
\bibitem{Tulin:2012uq} 
  S.~Tulin, H.~-B.~Yu and K.~M.~Zurek,
  %``Three Exceptions for Thermal Dark Matter with Enhanced Annihilation to $\gamma \gamma$,''
  arXiv:1208.0009 [hep-ph].
%
%%%%%%%%% Resonance models for Fermi gamma-ray line %%%%%%%%%
\bibitem{resonance}
%\cite{Lee:2012bq}
%\bibitem{Lee:2012bq}
  H.~M.~Lee, M.~Park and W.~-I.~Park,
  %``Fermi Gamma Ray Line at 130 GeV from Axion-Mediated Dark Matter,''
  Phys.\ Rev.\ D {\bf 86}, 103502 (2012);
%  [arXiv:1205.4675 [hep-ph]]; 
%\cite{Lee:2012wz}
%\bibitem{Lee:2012wz}
  H.~M.~Lee, M.~Park and W.~-I.~Park,
  %``Axion-mediated dark matter and Higgs diphoton signal,''
  JHEP {\bf 1212}, 037 (2012);
%  arXiv:1209.1955 [hep-ph];
  %\cite{Lee:2012ph}
%\bibitem{Lee:2012ph}
  H.~M.~Lee, M.~Park and V.~Sanz,
  %``Interplay between Fermi gamma-ray lines and collider searches,''
  JHEP {\bf 1303}, 052 (2013);
%  [arXiv:1212.5647 [hep-ph]].
  D.~Das, U.~Ellwanger and P.~Mitropoulos,
  %``A 130 GeV photon line from dark matter annihilation in the NMSSM,''
  JCAP {\bf 1208}, 003 (2012);
%  [arXiv:1206.2639 [hep-ph]].
%\cite{SchmidtHoberg:2012ip}
%\bibitem{SchmidtHoberg:2012ip}
  K.~Schmidt-Hoberg, F.~Staub and M.~W.~Winkler,
  %``Enhanced diphoton rates at Fermi and the LHC,''
  JHEP {\bf 1301}, 124 (2013).
 % [arXiv:1211.2835 [hep-ph]].
  %%CITATION = ARXIV:1211.2835;%%
  %23 citations counted in INSPIRE as of 04 Apr 2013



%
\bibitem{resonance2}
%\cite{Dudas:2009uq}
%\bibitem{Dudas:2009uq}
  E.~Dudas, Y.~Mambrini, S.~Pokorski and A.~Romagnoni,
  %``(In)visible Z-prime and dark matter,''
  JHEP {\bf 0908} (2009) 014
  [arXiv:0904.1745 [hep-ph]];
  %%CITATION = ARXIV:0904.1745;%%
  %45 citations counted in INSPIRE as of 04 Apr 2013
%\cite{Mambrini:2009ad}
%\bibitem{Mambrini:2009ad}
  Y.~Mambrini,
  %``A Clear Dark Matter gamma ray line generated by the Green-Schwarz mechanism,''
  JCAP {\bf 0912} (2009) 005
  [arXiv:0907.2918 [hep-ph]];
  %%CITATION = ARXIV:0907.2918;%%
  %44 citations counted in INSPIRE as of 04 Apr 2013
  %\cite{Dudas:2012pb}
%\bibitem{Dudas:2012pb}
  E.~Dudas, Y.~Mambrini, S.~Pokorski and A.~Romagnoni,
  %``Extra U(1) as natural source of a monochromatic gamma ray line,''
  JHEP {\bf 1210} (2012) 123
  [arXiv:1205.1520 [hep-ph]];
  %%CITATION = ARXIV:1205.1520;%%
  %53 citations counted in INSPIRE as of 04 Apr 2013
 %\cite{Jackson:2009kg}
%\bibitem{Jackson:2009kg}
  C.~B.~Jackson, G.~Servant, G.~Shaughnessy, T.~M.~P.~Tait and M.~Taoso,
  %``Higgs in Space!,''
  JCAP {\bf 1004} 004 (2010), 
  %[arXiv:0912.0004 [hep-ph]],
% C.~B.~Jackson, G.~Servant, G.~Shaughnessy, T.~M.~P.~Tait and M.~Taoso,
  %``Gamma-ray lines and One-Loop Continuum from s-channel Dark Matter Annihilations,''
  arXiv:1302.1802 [hep-ph],
%  C.~B.~Jackson, G.~Servant, G.~Shaughnessy, T.~M.~P.~Tait and M.~Taoso,
  %``Gamma Rays from Top-Mediated Dark Matter Annihilations,''
  arXiv:1303.4717 [hep-ph].


%%%%%%%%%%%  Vector Higgs portal  %%%%%%%%%%%%%%
\bibitem{vdm}
  S.~Kanemura, S.~Matsumoto, T.~Nabeshima and N.~Okada,
  %``Can WIMP Dark Matter overcome the Nightmare Scenario?,''
  Phys.\ Rev.\ D {\bf 82}, 055026 (2010).

\bibitem{vdm2}  
  O.~Lebedev, H.~M.~Lee and Y.~Mambrini,
  %``Vector Higgs-portal dark matter and the invisible Higgs,''
  Phys.\ Lett.\ B {\bf 707}, 570 (2012);
  A.~Djouadi, O.~Lebedev, Y.~Mambrini and J.~Quevillon,
  %``Implications of LHC searches for Higgs--portal dark matter,''
  Phys.\ Lett.\ B {\bf 709}, 65 (2012).


\bibitem{farzan} 
%\cite{Farzan:2012kk}
%\bibitem{Farzan:2012kk} 
  Y.~Farzan and A.~R.~Akbarieh,
  %``Natural explanation for 130 GeV photon line within vector boson dark matter model,''
  arXiv:1211.4685 [hep-ph].


%\cite{Abe:2012hb}
\bibitem{Abe:2012hb} 
  T.~Abe, M.~Kakizaki, S.~Matsumoto and O.~Seto,
  %``Vector WIMP Miracle,''
  Phys.\ Lett.\ B {\bf 713}, 211 (2012).

%  E.~Dudas, Y.~Mambrini, S.~Pokorski and A.~Romagnoni,
  %``Extra U(1) as natural source of a monochromatic gamma ray line,''
   %JHEP {\bf 1210}, 123 (2012);


\bibitem{pko}
%\cite{Baek:2012se}
%\bibitem{Baek:2012se}
  S.~Baek, P.~Ko, W.~-I.~Park and E.~Senaha,
  %``Higgs Portal Vector Dark Matter : Revisited,''
  arXiv:1212.2131 [hep-ph].
  %%CITATION = ARXIV:1212.2131;%%


\bibitem{triplet}
%\cite{Wang:2012ts}
%\bibitem{Wang:2012ts}
  L.~Wang and X.~-F.~Han,
  %``130 GeV gamma-ray line and enhancement of $h\to\gamma\gamma$ in the Higgs triplet model plus a scalar dark matter,''
  Phys.\ Rev.\ D {\bf 87}, 015015 (2013)
  [arXiv:1209.0376 [hep-ph]].
  %%CITATION = ARXIV:1209.0376;%%
  %31 citations counted in INSPIRE as of 04 Apr 2013   


%%%%%%%%% Triplet Higgs %%%%%%%%%%%%%%
\bibitem{tripletH}
%\cite{Chun:2012jw}
%\bibitem{Chun:2012jw}
  E.~J.~Chun, H.~M.~Lee and P.~Sharma,
  %``Vacuum Stability, Perturbativity, EWPD and Higgs-to-diphoton rate in Type II Seesaw Models,''
  JHEP {\bf 1211}, 106 (2012).

\bibitem{kannike}
  K.~Kannike,
  %``Vacuum Stability Conditions From Copositivity Criteria,''
  Eur.\ Phys.\ J.\ C {\bf 72}, 2093 (2012).

%%%%%%%%%%%%%%%%%%%%%%%%%%%%%%%%%%%%%%%%%%%%%%%%%%%%%%
\bibitem{Boltzmann}
%\bibitem{KolbTurner}
E.~W.~Kolb and M.~S.~Turner, {\it The Early Universe},Addison-Wesley (1990).
%
\bibitem{Gondolo:1990dk}
P.~Gondolo and G.~Gelmini,
%``Cosmic abundances of stable particles: Improved analysis,''
Nucl.\ Phys.\  B {\bf 360}, 145 (1991).

%%%%%%%%% Planck %%%%%%%%%%%%%%%%%%%%%%%%%
%\cite{Ade:2013lta}
\bibitem{Ade:2013lta}
  P.~A.~R.~Ade {\it et al.}  [ Planck Collaboration],
  %``Planck 2013 results. XVI. Cosmological parameters,''
  arXiv:1303.5076 [astro-ph.CO].

%%%%%%%%  Continuum constraints  %%%%%%%%%%%%%%%%%
%\cite{Buchmuller:2012rc}
\bibitem{Buchmuller:2012rc} 
  W.~Buchmuller and M.~Garny,
  %``Decaying vs Annihilating Dark Matter in Light of a Tentative Gamma-Ray Line,''
  JCAP {\bf 1208}, 035 (2012).
\bibitem{Cohen:2012me} 
  T.~Cohen, M.~Lisanti, T.~R.~Slatyer and J.~G.~Wacker,
  %``Illuminating the 130 GeV Gamma Line with Continuum Photons,''
  JHEP {\bf 1210}, 134 (2012).
\bibitem{Asano:2012zv} 
  M.~Asano, T.~Bringmann, G.~Sigl and M.~Vollmann,
  %``The 130 GeV gamma-ray line and generic dark matter model building 
  %constraints from continuum gamma rays, radio and antiproton data,''
  arXiv:1211.6739 [hep-ph].


%\cite{Griest:1990kh}
\bibitem{Griest:1990kh}
  K.~Griest and D.~Seckel,
  %``Three exceptions in the calculation of relic abundances,''
  Phys.\ Rev.\ D {\bf 43}, 3191 (1991).
  %%CITATION = PHRVA,D43,3191;%%
  %523 citations counted in INSPIRE as of 02 Apr 2013


\bibitem{jijifan}
%\cite{Fan:2013qn}
%\bibitem{Fan:2013qn}
  J.~Fan and M.~Reece,
  %``Probing Charged Matter Through Higgs Diphoton Decay, Gamma Ray Lines, and EDMs,''
  arXiv:1301.2597 [hep-ph].
  %%CITATION = ARXIV:1301.2597;%%
  %9 citations counted in INSPIRE as of 04 Apr 2013



%%%%%%%%  Higgs diphoton: experiment  %%%%%%%%%%%
\bibitem{March14}
ATLAS-CONF-2013-012; CMS-HIG-13-001.


%%%%%%%%  Higgs diphoton: combining data  %%%%%%%
\bibitem{Baglio:2012et} 
J.~Baglio, A.~Djouadi and R.~M.~Godbole,
  %``The apparent excess in the Higgs to di-photon rate at 
  %the LHC: New Physics or QCD uncertainties?,''
  Phys.\ Lett.\ B {\bf 716}, 203 (2012).
\bibitem{Espinosa:2012im} 
  J.~R.~Espinosa, C.~Grojean, M.~Muhlleitner, and M.~Trott,
  %``First Glimpses at Higgs' face,''
  JHEP {\bf 1212}, 045 (2012).

%%%%%%%%  Higgs diphoton: formula  %%%%%%%%%%%%%%
\bibitem{HHG}
J.~F.~Gunion, H.~E.~Haber, G.~Kane, and S.~Dawson,
 {\it The Higgs Hunter's Guide}, Addison-Wesley (1990).

%\bibitem{LHCtracker}
%ATLAS-CONF-2012-075; CMS-PAS-EXO-12-026.

\bibitem{RPV}
%\cite{Allanach:1999ic}
%\bibitem{Allanach:1999ic}
  B.~C.~Allanach, A.~Dedes, and H.~K.~Dreiner ,
  %``Bounds on R-parity violating couplings at the weak scale and at the GUT scale,''
  Phys.\ Rev.\ D {\bf 60} 075014 (1999);
F.~Ledroit and G.~Sajot, GDR-S-008. 

  
%%%%%%%%% mu-e gamma and g-2 from LLS coupling  %%%%%%%%%%
%\cite{Dicus:2001ph}
\bibitem{Dicus:2001ph} 
  D.~A.~Dicus, H.~-J.~He and J.~N.~Ng,
  %``Neutrino - lepton masses, Zee scalars and muon g-2,''
  Phys.\ Rev.\ Lett.\  {\bf 87}, 111803 (2001);
%  [hep-ph/0103126].
  A.~Ghosal, Y.~Koide and H.~Fusaoka,
  %``Lepton flavor violating Z decays in the Zee model,''
  Phys.\ Rev.\ D {\bf 64}, 053012 (2001),
%  [hep-ph/0104104].
  C.~A.~De Sousa Pires and P.~S.~Rodrigues da Silva,
  %``Electric charge quantization and the muon anomalous magnetic moment,''
  Phys.\ Rev.\ D {\bf 65}, 076011 (2002);
%  [hep-ph/0108200].
  K.~Cheung and O.~Seto,
  %``Phenomenology of TeV right-handed neutrino and the dark matter model,''
  Phys.\ Rev.\ D {\bf 69}, 113009 (2004).
%  [hep-ph/0403003].
 
 
\bibitem{barger}
%\cite{Barger:1989rk}
%\bibitem{Barger:1989rk} 
  V.~D.~Barger, G.~F.~Giudice and T.~Han,
  %``Some New Aspects of Supersymmetry R-Parity Violating Interactions,''
  Phys.\ Rev.\ D {\bf 40}, 2987 (1989).
  %%CITATION = PHRVA,D40,2987;%%
  %529 citations counted in INSPIRE as of 20 May 2013
 
  
%%%%%%%%%  g-2 from LLS coupling  %%%%%%%%%%
%\cite{Kim:2001se}
\bibitem{Kim:2001se}
  J.~E.~Kim, B.~Kyae and H.~M.~Lee,
  %``Effective supersymmetric theory and (g-2)(muon with R-parity violation,''
  Phys.\ Lett.\ B {\bf 520}, 298 (2001).
%  [hep-ph/0103054].

%\cite{Nebot:2007bc}
\bibitem{Nebot:2007bc}
  M.~Nebot, J.~F.~Oliver, D.~Palao and A.~Santamaria,
  %``Prospects for the Zee-Babu Model at the CERN LHC and low energy experiments,''
  Phys.\ Rev.\ D {\bf 77}, 093013 (2008).
%  [arXiv:0711.0483 [hep-ph]].



\bibitem{ATLASslepton}  
ATLAS collaboration,   ATLAS-CONF-2012-076.
  
  
\bibitem{CMSslepton}  
CMS collaboration, CMS PAS SUS-12-022. 
  
  
\bibitem{LEPslepton}  
  A.~Heister {\it et al.}  [ALEPH Collaboration],
  %``Search for scalar leptons in e+ e- collisions at center-of-mass energies up to 209-GeV,''
  Phys.\ Lett.\ B {\bf 526}, 206 (2002).


\bibitem{Higgsfit}
%\cite{Falkowski:2013dza}
%\bibitem{Falkowski:2013dza}
  A.~Falkowski, F.~Riva and A.~Urbano,
  %``Higgs At Last,''
  arXiv:1303.1812 [hep-ph];
%\cite{Giardino:2013bma}
%\bibitem{Giardino:2013bma}
  P.~P.~Giardino, K.~Kannike, I.~Masina, M.~Raidal and A.~Strumia,
  %``The universal Higgs fit,''
  arXiv:1303.3570 [hep-ph].


\bibitem{VSB}
%\cite{Degrassi:2012ry}
%\bibitem{Degrassi:2012ry}
  G.~Degrassi, S.~Di Vita, J.~Elias-Miro, J.~R.~Espinosa, G.~F.~Giudice, G.~Isidori and A.~Strumia,
  %``Higgs mass and vacuum stability in the Standard Model at NNLO,''
  JHEP {\bf 1208}, 098 (2012).
%  [arXiv:1205.6497 [hep-ph]].

\bibitem{giudice}  
  J.~Elias-Miro, J.~R.~Espinosa, G.~F.~Giudice, H.~M.~Lee and A.~Strumia,
  %``Stabilization of the Electroweak Vacuum by a Scalar Threshold Effect,''
  JHEP {\bf 1206}, 031 (2012).
%  [arXiv:1203.0237 [hep-ph]].
  
\bibitem{lebedev}  
  O.~Lebedev,
  %``On Stability of the Electroweak Vacuum and the Higgs Portal,''
  Eur.\ Phys.\ J.\ C {\bf 72}, 2058 (2012).
%  [arXiv:1203.0156 [hep-ph]].
 
\bibitem{batell}  
  B.~Batell, S.~Jung and H.~M.~Lee,
  %``Singlet Assisted Vacuum Stability and the Higgs to Diphoton Rate,''
  JHEP {\bf 1301}, 135 (2013).
%  [arXiv:1211.2449 [hep-ph]].


\end{thebibliography}
\end{document}